\documentclass[11pt,twoside]{article}
\usepackage{acta-info}
\usepackage{euler}
\usepackage{rotating}
\usepackage{multirow}

\newcommand{\vol}{\textbf{3,} 2 (2011) 158--171}   
\newcommand{\amss}{\amssubj{%
05A10, 11Y11
}}     
\newcommand{\CRs}{\CRsubj{%
G.4
}}   
\newcommand{\keyw}{\keywords{%
generalized Pascal triangles, elliptic curve primality proving
}}

\newfont{\ms}{msbm10 scaled\magstep1}
\newcommand{\QQ}{\mbox{\ms Q}}

\newcommand{\ZZ}{\mbox{\ms Z}}

\newcommand{\alge}{%
\begin{tabbing}%
99 \= xxx\=xxx\=xxx\=xxx\=xxx\=xxx\=xxx\=xxx \+ \kill
}

\newcommand{\algeN}{%
\settowidth{\temp}{199}
\setlength{\temp}{6mm-\temp}
\begin{tabbing}%
\hspace*{1pt}999\=xxx\=xxx\=xxx\=xxx\=xxx\=xxx\=xxx\=xxx \+ \kill  
}

\newcommand{\algv}{%
\end{tabbing}
}

\newenvironment{alg}[1]{%
\vspace{4mm}  
\vbox\bgroup\noindent\textsc{#1}%
\vspace*{-1mm}  
\alge}
{
\algv\egroup
\vspace{0mm}  
}

\def\algnewpage
{
\algv\egroup
\newpage
\vbox\bgroup
\alge
}

\def\algnewpageN
{
\algv\egroup
\newpage
\vbox\bgroup
\algeN
}

\newcommand{\algkoz}[1]{
\algv\egroup
\pushtabs
\vspace{-11mm}
\vspace{#1}
\vbox\bgroup
\alge
\poptabs}

\newcommand{\key}[1]{\textbf{#1}}

\newcommand{\ideze}{\setbox0=\hbox{\lower1.38ex\hbox{''}}\dp0=0pt\box0}
\newcommand{\abs}[1]{\left| #1\right|}

\begin{document}

\title{%
Large primes in generalized Pascal triangles
}

\maketitle

\oneauthor{%
\href{http://compalg.inf.elte.hu}{G\'abor FARKAS}
}{%
\href{http://www.elte.hu}{E\"otv\"os Lor\'and University}
}{%
 \href{mailto:farkasg@compalg.inf.elte.hu}{farkasg@compalg.inf.elte.hu}
}


\twoauthors{%
\href{http://www.sze.hu/~kallos}{G\'abor KALL\'OS}
}{%
\href{http://www.sze.hu}{Sz\'echenyi Istv\'an University, Gy\H or}
}{%
\href{mailto:kallos@sze.hu}{kallos@sze.hu}
}{%
\href{http://compalg.inf.elte.hu}{Gy\"ongyv\'er KISS}
}{%
\href{http://www.elte.hu}{E\"otv\"os Lor\'and University}
}{%
\href{mailto:kissgyongyver@gmail.com}{kissgyongyver@gmail.com}
}


\short{%
G. Farkas, G. Kall\'os, G. Kiss
}{%
Large primes in generalized Pascal triangles
}

\begin{abstract}
In this paper, after presenting the results of the
generalization of Pascal triangle (using powers of base numbers),
we examine some properties of the $112$-based triangle, most of
all regarding to prime numbers. Additionally, an effective
implementation of ECPP method is presented which enables Magma
computer algebra system to prove the primality of numbers with
more than $1000$ decimal digits.
\end{abstract}


\section{Generalized Pascal triangles using the powers of base numbers}

As it is a well-known fact, the classic Pascal triangle has served
as a model for various generalizations. Among the broad variety of
ideas of generalizations we can find e.g.: the generalized
binomial coefficients of $s^\mathrm{th}$ order (leading to
generalized Pascal triangles of $s^\mathrm{th}$ order), the
multinomial coefficients (leading to Pascal pyramids and
hyperpyramids), special arithmetical sequences (leading to
resulting triangles which we might call as Lucas, Fibonacci,
Gaussian, Catalan, ... triangle) (details in \cite{bo}).

One of the present authors has devised, and then worked out in detail and published such a type of generalization, which is
based on the idea of using ``the powers of the base number". Referring
to our former results (presented in detail in \cite{ka1} and
\cite{ka2}; here we don't repeat/echo the theorems and
propositions) we show here the first few rows of the 112-based
triangle (Figure \ref{fig:tri112}), which will gain outstanding
importance below in this paper.

\begin{figure}[h]
\begin{center}
\begin{tabular}{c@{\hspace{4mm}}c@{\hspace{4mm}}c@{\hspace{4mm}}
c@{\hspace{3mm}}c@{\hspace{3mm}}c@{\hspace{2mm}}c@{\hspace{2mm}}
c@{\hspace{2mm}}c@{\hspace{2mm}}c@{\hspace{2mm}}c@{\hspace{2mm}}
c@{\hspace{2mm}}c@{\hspace{2mm}}c@{\hspace{2mm}}c@{\hspace{2mm}}
c@{\hspace{2mm}}c@{\hspace{2mm}}c@{\hspace{2mm}}c@{\hspace{2mm}}
c@{\hspace{2mm}}c@{\hspace{2mm}}c@{\hspace{2mm}}c@{\hspace{2mm}}c@{\hspace{2mm}}c}
& & & & & & & & & & & & 1 & & & & & & & & & & & &\\
& & & & & & & & & & 1 & & 1 & & 2 & & & & & & & & & &\\
& & & & & & & & 1 & & 2 & & 5 & & 4 & & 4 & & & & & & & &\\
& & & & & & 1 & & 3 & & 9 & & 13 & & 18 & & 12 & & 8 & & & & & &\\
& & & & 1 & & 4 & & 14 & & 28 & & 49 & & 56 & & 56 & & 32 & & 16 & & & &\\
& & 1 & & 5 & & 20 & & 50 & & 105 & & 161 & & 210 & & 200 & & 160 & & 80 & & 32 & &\\
1 & & 6 & & 27 & & 80 & & 195 & & 366 & & 581 & & 732 & & 780 & &
640 & & 432 & & 192 & & 64\\
& & & & & & & & & & & & \dots & & & & & & & & & & & &\\
\end{tabular}\\
\end{center}
\caption{The 112-based triangle}\label{fig:tri112}
\end{figure}

Let us use the notation $E^{a_0 a_1 \dots a_{m-1}}_{k,n}$ for the
$k^\mathrm{th}$ element in the $n^\mathrm{th}$ row of $a_0 a_1
\dots a_{m-1}$-based triangle ($0\le a_0,a_1,\dots ,a_{m-1}\le 9$
are integers). Then we have the definition rule, as follows:
$$E^{a_0 a_1 \dots a_{m-1}}_{k,n} = a_{m-1}E^{a_0 a_1 \dots
a_{m-1}}_{k-m+1,n-1} + a_{m-2}E^{a_0 a_1 \dots a_{m-1}}_{k-m+2,n-1}
+\cdots + $$
$$+a_1E^{a_0 a_1 \dots a_{m-1}}_{k-1,n-1} + a_0E^{a_0 a_1
\dots a_{m-1}}_{k,n-1}.$$ The indices in the rows and columns run
from 0, elements with non-existing indices are considered to be
zero. Applying this general form to the $112$-based triangle
(now: $m=3$), we get the specific rule
$$E^{112}_{k,n} = 2 E^{112}_{k-2,n-1} + E^{112}_{k-1,n-1}
+ E^{112}_{k,n-1}.$$

The historical overview of this special field is presented in
\cite{ka2}. In the last few years there were published several new
results which are related to our topic (e.g. \cite{be}). Moreover,
besides that, up to about 2005, all generalized
triangle sequences of the type $ax + by$ were added to the
database On-line Encyclopedia of Integer sequences \cite{sl},
since that time there have been several new applications, too, based on
sequences appearing in our triangles. However, e.g. the sequences
based on the general $abc$-based triangles are still not
widely known.

Recalling the basic properties of generalized
triangles---most of all in connection with powering the base
number $a_0 a_1 \dots a_{m-1}$ and with the polynomial $(a_0 x_0 +
a_1 x_1 + \cdots + a_{m-1} x_{m-1})^n$---we can state that
we have the ``right" to call these types of triangles as generalized
Pascal triangles (details in \cite{ka2}, summary in \cite{fa-ka}).

\section{Divisibility of elements and prime numbers}

The classic divisibility investigations in Pascal triangle (for
binomial coefficients) are very popular and even spectacular, if the
traditional ``strict" mathematical approach is moved toward
coloring and fractals (details in \cite{bo}). For generalized
binomial coefficients (with our notation: in triangles with bases
$11 \cdots 1$) we have similar results, too, with a remark that in
these cases general proofs are harder, and there are many
conjectures, too.

We recall here the beautiful result of Richard C.
Bollinger, who proved for generalized Pascal triangles of
$p^\mathrm{th}$ order that for large $n$, ``almost every" element
in the $n^\mathrm{th}$ row is divisible by $p$ (see \cite{bo}, p. 24). For example, for
the $111$-based triangle this means divisibility by $3$. (We mention that 
the $p^\mathrm{th}$ order Pascal triangle is a triangle with base 
$11 \cdots 1$, where we have $p$ pieces of $1$.)

Now we turn our attention specially to the $112$-based triangle,
and in the following we are interested mostly in prime numbers. It
is obvious that the right part of the triangle contains only even
numbers. Moreover, if we move to the right, the powers of $2$ are
usually (not always) growing as divisors. Analyzing connections
with the multinomial theorem we can conclude that the left part of
the triangle contains mostly (with possible exception of the first
two places) composite numbers, too. Of course, this can be not true
for the $0^\mathrm{th}$ and $1^\mathrm{st}$ numbers, which are the
same as in the classic Pascal triangle. Moreover, using induction
we can see that the center element in every row is always odd.

We can pose obviously two (not hard) questions in connection with
prime numbers:

1. Can we find every prime number as an element in our triangle?

2. Can we find every prime number as an element in our triangle in
non-trivial places?

The answer to question 1 is ``yes", as we already saw above (the
$1^\mathrm{st}$ elements in every row, however, this is a trivial
match). To question 2, we fix first that
primes are worth looking for only in the middle position.

With a computer investigation (using e.g. the Maple program) we
can find 6 small primes up to the $100^\mathrm{th}$
row (Figure \ref{fig:tri112a}).

\begin{figure}[h]
\begin{center}
\begin{tabular}{l@{\hspace{4mm}}r@{\hspace{4mm}}c}
Position (row, column) & Prime\\
2, 2 & 5\\
3, 3 & 13\\
8, 8 & 7393\\
15, 15 & 65753693\\
21, 21 & 175669746209\\
24, 24 & 9232029156001\\
\end{tabular}\\
\end{center}
\caption{Small primes in the 112-based triangle}
\label{fig:tri112a}
\end{figure}

Extending the examination up to the $1900^\mathrm{th}$ row, we get
only one more positive answer, in position $(156, 156)$, a
$90$-digit prime (candidate). So, the answer to our second
question (considering only this triangle) is ``no".

Our possibilities are extended rapidly, if we look up not only
pure prime numbers, but even decompositions. So now we modify our
question 2 as ``can we find every prime number as a factor of any
element in our triangle?" (Examining only non-trivial places, so,
positions 0 and 1 are in every row excluded.)

We see immediately that every one-digit prime occurs as a factor
at least once up to the $4^\mathrm{th}$ row. Here 2 and 5 are
triangle elements themselves; 3 is a factor of 9, 7 is a factor of
14.

Continuing with an easy computer examination for two-digit primes
we find all but $4$ up to the $12^\mathrm{th}$ row. For the rest
of the numbers we get the following first occurrences (in number--row form): 79--14, 71--15, 59--17 and, surprisingly 41--27.

Now, we turn our attention to 3-digit primes. Here we need a much
larger triangle-part. Let's choose, say, a 100-row triangle in an
easy-factorized form. With a small Maple program on a normal
table-PC, we can generate the necessary data in a few minutes. (Easy
factorization is very important here, otherwise, with full
factorization the generation could take an extremely long time...) The output of the program in txt form
will be approximately 1.15 MBytes.

From the 143 3-digit primes we find 105 up to $40^\mathrm{th}$
row. For the remaining 38 numbers, 18 numbers are situated in rows $41-50$, 11
additional primes in rows $51-60$, and 2 (823 and 827) in rows
$61-70$. The still missing ``hardest" 3-digit primes finally give
the following first occurrences (in number -- row form): 479 --
74, 499 -- 74, 677 -- 76, 719 -- 77, 859 -- 72, 937 -- 98 and 947
-- 73. To the contrary, the ``easiest" 3-digit primes are 103, 191
and 409 in the $7^\mathrm{th}$ row.

With this we give up the claim ``to find all of the primes as
divisors".

Our next investigation focuses on very large prime
factors (more accurately: prime candidates).





Computer investigations suggest that the largest prime factors in a
given row occur very likely in the center position or very close
to that place. Of course, this is not an absolute rule, but since
our goal is ``only" to find very large prime (candidate) factors,
we can limit the investigation to the center element. (This has a
significant importance to achieving: go as ``deep" relatively quickly
in the triangle as possible.)

Moreover, the center element carries special properties compared
with other elements. Recalling Richard C. Bollinger's
result above, we can set up a similar interesting conjecture:

\emph{For large $n$, the center element in the $n^\mathrm{th}$ row
``almost surely" will be divisible by $5$ and $7$ (but surely not
by $2$ and usually not by $3$).}


So, with a relatively simple Maple program  we set out to
the easy-factorization of the center element up to the 
$1900^\mathrm{th}$ row. On a normal table-PC, the execution time
is approximately 11 hours, with an output file in txt form roughly
110 KBytes.

Analyzing the output we can deduce that prime divisors here follow
the Knuth-observation \cite{kn}, too: we usually find few small
factors some of which are repetitive; composite (not decomposable
with the 'easy' option) large factors are common, pure large prime
factors are however rare or extremely rare.

\begin{figure}[h]
\begin{center}
\begin{tabular}{l@{\hspace{4mm}}r@{\hspace{4mm}}c}
Position (row, column) & Digits of the prime candidate\\
1726, 1726 & 1002\\
1793, 1793 & 1028\\
1794, 1794 & 1030\\
\end{tabular}\\
\end{center}
\caption{Large ``pure" prime factors (candidates)---112-based triangle, center position} \label{fig:abra3}
\end{figure}

Considering only the primes (prime candidates) with digits more
than 1000 we get 3 matches.

Here the second and third matches are especially interesting, since
they can be considered as a special kind of ``twin-primes"
(candidates) in the triangle. In general, our chance to find
``pure" large prime factors in consecutive rows is very little...

Here the factorization of element with position $1793, 1793$ is as follows:
\newpage

\vspace{1mm}
\noindent \ \ $1793, 1793;``(5)*``(7)^2*``(673)*``(65119)*``(1485703)*``(15578887875328\\
926423851777567602680378792003694589981499750631818308971422277975\\
902867850432471811687112334064063828539296067422531997963055491323\\
406425659317001574425151788919713654021679547897110675223861482309\\
644220358490739245691930715715021145166205571510978302005857149111\\
239471032734380710285002174983967604232152940389858538629493812650\\
108566716591594874813194189360195173091031608755605756723631900973\\
625032697091409833078265261680211635427069757196618031458397872466\\
034789488450265204214587550269112317436588892430166513888148357222\\
480962630168478230243146450158020142586939406221546644931686618139\\
068737541801842683626194613956159330873776421795220707554672321055\\
658602305273678940456712151943459348907356567358277310497505925970\\
210070347980231047308886323693790450859256057748541430119354204022\\
527748661261790305800487349106563678280226712828838174678186252307\\
070941149885645163684441661612796581751766644659424590726902531393\\
104098376100305217952214533052008783687240950373043230661705142861\\
901235736247002277563333)$

\vspace{1mm}

In \cite{fa-ka} we proved the primality of the largest factor of $1726, 1726$ which
has $1002$ decimal digits. That time we used a freeware software developed
by F. Morain.
In the remaining part of this paper our goal is to present our selfmade
program which is appropriate to prove the prime property of such large
numbers. Let us denote the $1028$ digits long factor of $1793, 1793$ by $n_1$ and
the $1030$ digits long factor of $1794, 1794$ by $n_2$. We investigated  $n_1$ and $n_2$
with our program, and have found that both of them are really primes. Moreover,
the process of the proof and shematic structure of the evidence will be presented, too.

\section{Atkin's primality test}

We described the theoretical foundations of the elliptic curve
primality proving in \cite{fa-ka}. Unfortunately, most computer
algebra systems include just probability primality test, so we can
not use them to reach our purposes. Although the Magma system
(described below) is able to carry out primality proving with
ECPP (Elliptic Curve Primality Proving), we did not get any result even
after two days running for $n_2$. Thus we have
developed an own primality proving program presented in the next
section.

According to the notation of
\cite{fa-ka} let us denote an elliptic curve over $\ZZ/n\ZZ$ by
$E_n$. The first step in the basic ECPP algorithm is choosing
randomly an $E_n$ elliptic curve, the second one is counting
$|E_n|$, the order of $E_n$. The latter action is very time-consuming, so
we had to find an improved version of ECPP. Finally we have implemented an algorithm suggested by A. O. L. Atkin. A specification of
this method can be found in \cite{at-mo}.  Lenstra and Lenstra published a
heuristic running time analysis of Atkin's elliptic curve primality
proving algorithm in \cite{ll}. They conjectured that with fast arithmetic
methods the running time of ECPP can be reduced to
$O(\ln^{4+\epsilon}(n))$.

Atkin brilliant idea was founding an appropriate $m$
order in advance and then constructing $E_n$ for this $m$ avoiding
the order-counting. Moreover, we get simultaneously two elliptic
curves increasing the chance of the successful running of the
test. $m$ order has to be chosen from the algebraic integer of an
imaginary quadratic field $\QQ(\sqrt{D})$. An appropriate $D$,
so-called {\it fundamental discriminant}, has some properties:
$D\equiv 0\pmod{4}$, or $D\equiv 1\pmod{4}$, for every $k(>1)$
$D/k^2$ is not a fundamental discriminant, $D \leq -7$ and
$(D|n)=1$, where $(D|n)$ is the Jacobi symbol.

The function \textsc{NextD}$()$ gives a value $D$ which meets the above mentioned requirements. A given
$D$ value is suitable if there exist such $x,y \in \ZZ$ for which

\begin{equation}\label{egy_1}
4n=(2x+yD)^2-y^2D.
\end{equation}
In that case we get two possible orders: $m=\abs{\nu\pm 1}^2$,
where
$$\nu = x + y\frac{D+\sqrt{D}}{2}.$$

If $(\ref{egy_1})$ is valid, then we can compute an $x_0$ root of
the {\it Hilbert polynomial} $\pmod{n}$. The function
\textsc{Hilbert}$(n,D)$ returns with a root of the appropriate
Hilbert polynomial. Then we get two elliptic curves with order
$m=\abs{\nu\pm 1}^2$. The rest of the algorithm works as we
described in \cite{fa-ka}.

\begin{alg}{Proof($E_n,m,f$)}
1 \' $P \leftarrow$ \textsc{Randompoint}($E_n$)\\
2 \' \key{if} \= $f\cdot P$ is not defined\\
3 \'             \> \key{then} \= \key{return}\mbox{ {\sc composite}}\\
4 \' \key{if} \= $f\cdot P=O$\\
5 \'             \> \key{then} \= \key{goto}\mbox{ 1}\\
6 \' \key{if} \= $mP\neq O$\\
7 \'             \> \key{then} \= \key{return}\mbox{ {\sc no}}\\
8 \' \key{return}\mbox{ {\sc yes}}
\end{alg}

Here symbol $O$ means the ``point infinitely far" e.g. the unit of the Abelian group. 
The function \textsc{Proof}$()$ has three input values: $E_n$,
$m$, $f$, where $E_n$ is an elliptic curve with order $m$,
$m=f\cdot s$, the factorization of $f$ is known and $s$ is
probably prime. The output value \textsc{composite} means that $n$
is surely composite. If the output is \textsc{no}, then $n$ is
composite or we have to choose the other elliptic curve. In case
\textsc{yes} the next recursion step follows. In the following we
present the pseudocode of the Atkin's test.

\begin{alg}{Atkin-primality-test($n$)}
1 \' $D \leftarrow$ \textsc{NextD}$()$\\
2 \' $\omega \leftarrow (D+\sqrt{D})/2$\\
3 \' \key{if} \= $\exists \ x,y\in \ZZ : 4n=(2x+yD)^2-y^2D$\\
4 \'             \> \key{then} \= $\nu \leftarrow x+y\omega$ \\
5 \'             \> \key{else} \> \key{goto}\mbox{ 1}\\
6 \' $m \leftarrow \abs{\nu+1}^2$\\
7 \' \key{if} \= $m=f\cdot s$, where $s$ ``probably prime" and $s> \left( \sqrt[4]{n}+1\right)^2$\\
8 \'             \> \key{then} \= \key{goto}\mbox{ 12}\\
9 \' $m \leftarrow \abs{\nu-1}^2$\\
10 \' \key{if} \= $m=f\cdot s$ can not be produced so that $s$ is ``probably prime"\\
   \'             \>            \>          \>         \>          \>and $s> \left( \sqrt[4]{n}+1\right)^2$\\
11 \'             \> \key{then} \= \key{goto}\mbox{ 1}\\
12 \' $x_0 \leftarrow$ \textsc{Hilbert}($n,D$)\\
13 \' $c\leftarrow$ arbitrary integer for which $(c/n)=-1$\\
14 \' $k\leftarrow$ arbitrary integer for which $k\equiv x_0/(1728-x_0)\pmod{n}$\\
15 \' $E_n\leftarrow \{ (x,y)\mid y^2=x^3+3kx+2k \}$\\
16 \' \key{if} \= \textsc{Proof}($E_n,m,f$) = \textsc{composite}\\
17 \'             \> \key{then} \= \key{return}\mbox{ {\sc composite}}\\
18 \'             \> \key{else} \= \key{if} \= \textsc{Proof}($E_n,m,f$) = \textsc{yes}\\
19 \'             \>            \>          \> \key{then} \= \key{goto}\mbox{ 23}\\
20 \'             \>            \> $E_n\leftarrow \{ (x,y)\mid y^2=x^3+3kc^2x+2kc^3 \}$ \\
21 \' \key{if} \= \textsc{Proof}($E_n,m,f$) = \textsc{composite} or \textsc{Proof}($E_n,m,f$) = \textsc{no} \\
22 \'             \> \key{then} \= \key{return}\mbox{ {\sc composite}}\\
23 \' \key{if} \= $s$ surely prime\\
24 \'             \> \key{then} \= \key{return}\mbox{ {\sc prime}}\\
25 \'             \> \key{else} \> \textsc{Atkin-primality-test($s$)}
\end{alg}

\section{Magma Computer Algebra System}
Magma \cite{magma} is a large software system specialized in
high-performance computations in number theory, group theory,
geometry, combinatorics and other branches of algebra. It was
launched  at the First Magma Conference on Computational Algebra
held at Queen Mary and Westfield College, London, August 1993. It
contains a large body of intrinsic functions (implemented in C
language), but also allows the user to implement functions on 
top of this, making use of the Pascal-like user language and the
programming environment that is provided.

\subsection{Primality tests in Magma}
Magma has several built-in functions for primality testing purposes. \vspace{2 mm}

{\tt IsProbablyPrime(}$n${\tt : {\sl parameter}) : RngIntElt }$\mapsto$ {\tt BoolElt}
\vspace{2mm}
 \newline The function returns {\sf TRUE} if and only if $n$ is a probable prime. This function uses the Miller-Rabin test; setting the optional integer
parameter {\tt Bases} to some value $B$, the Miller-Rabin test will use $B$ bases while testing compositeness. The default value is 20. This function will never
declare a prime number composite, but with very small probability (much smaller than $2^{-B}$, and by default less than $10^{-6}$) it may fail
to find a witness for compositeness, and declare a composite
number probably prime.\vspace{2 mm}

{\tt IsPrime(}$n${\tt : {\sl parameter}) : RngIntElt }$\mapsto$ {\tt BoolElt}
\vspace{2mm}
\newline This function proves primality using ECPP which is of course more time-consuming. It is possible though to set the optional Boolean
parameter {\tt Proof} to {\sf FALSE}; in which case the function
uses the probabilistic Miller-Rabin test, with the default number
of bases. \vspace{2 mm}

{\tt PrimalityCertificate(}$n${\tt : {\sl parameter}) : RngIntElt }$\mapsto$ {\tt List}
\vspace{2mm}
\newline This function proves primality and provides a certificate for it using ECPP. If the number $n$ is proven to be composite or the test fails, a runtime error occurs. \vspace{2 mm}

{\tt IsPrimeCertificate(}$c${\tt  : {\sl parameter}) : List }$\mapsto$ BoolElt
\vspace{2mm}
\newline To verify primality from a given certificate $c$ this function is used. This returns
the result of the verification by default, a more detailed outcome can be obtained by setting
the optional Boolean parameter {\tt ShowCertificate} to {\sf TRUE}. \vspace{3 mm}

The numbers $n_1$ and $n_2$ were tested with Magma's own ECPP,
using the intrinsic {\tt IsPrime} function, and with our ECPP
implementation written in Magma language. We refer to Magma's ECPP
algorithm as Magma-ECPP and to our implementation as
modified-ECPP. Both tests were running in Magma 2.16 on a machine
with 7425 MB RAM and four 2400 MHz Dual-Core AMD Opteron ({\sf
TM}) Processors.
\vspace{3 mm}

The Magma-ECPP provided a primality proof for $n_1$ in 32763.52 seconds, but seemed to stuck after the third iteration during the test of $n_2$; the modified-ECPP provided proof for $n_1$ in 5666.96 seconds and for $n_2$ in 5153.37 seconds. As the modified-ECPP is not finished yet, the running time can still be improved.

\subsection{The implementation of ECPP algorithm}

The ECPP algorithm consists of iteration steps, where the $i^{th}$
iteration step outputs an $s_i$ which will be the input of the
next iteration step. In one iteration step an attempt is made to
factor order $m_i$ of the group of points on a curve $E_i$. Curve
$E_i$ is defined using the input $s_{i-1}$ and  a discriminant of
an imaginary quadratic field, read in from a list.

If the attempt is successful, factor $s_i$ is the output; if not,
we need to backtrack. A different discriminant in an iteration
step results in a different $s_i$. The possible iteration chains
that occur this way, can be represented as paths in a directed
graph $G(n)$. The nodes of $G(n)$ are the $s_i$'s, the root
represents $n$, the edges are the iteration steps. An edge leads
from $s_i$ to $s_{i+1}$ if there is an iteration that produces
$s_{i+1}$ with input $s_i$. Consider a path successful if the
corresponding iteration-chain starts with input $n$ and ends
with input $s_l$, where $s_l$ is a small prime, which can be
verified by easy inspection, or trial division. In the rest of the
paper we refer to the $s_i$'s also as nodes.

Magma-ECPP uses a small fixed set of discriminants during the
process. Each iteration goes through this set until it finds a
discriminant which produces a new node. Using a small set of
discriminants makes the algorithm faster, but increases the
probability of producing no new node. If no discriminant produces
new node in the set it backtracks to the previous node and retries
that with the same set of discriminants but possibly stronger
factorization methods to factor the $m_i$'s. If backtracking does
not produce a new node, it will try to factor \pagebreak 
 again with more effort; these hard factorizations may consume a large
amount of time, and the process appears to get stuck in a seemingly endless
loop.
This happened during the test of our number $n_2$ with Magma-ECPP.

\subsubsection{Modifications}
During the iteration steps certain limits are used; for example,
the bound $B$ on the primes found in factoring the $m_i$-s.
Imposing a small $B$ decreases the difference between the size of
the $s_i$-s and thus may extend the path down to the small primes.
On the other hand, setting a large $B$ significantly increases the
running time needed for factoring. Of course, choosing a more
sophisticated factoring method smoothes the differences in running
time, but the size of $B$ still remains an important factor. Other
important limits are the bound $D$ on the discriminants and the
limit $S$ on the prime factors of the discriminants. Decreasing
them leads to speed improvement but to a smaller set of
discriminants, too.

The modified-ECPP uses a huge file which contains a list of fully factored discriminants up to $10^9$. During the selection of discriminants useful for the current input we extract a modular square root of its prime divisors and build up the square root of the discriminant by multiplication. After using one prime, the square root is stored, and thus it will be computed only once in an iteration step. The speed that we gain this way makes it possible to increase limits $D$, $S$ in the iterations, which are adjusted to the size of the current input.

The steps can be extended to result in a {\it series} of $s_i$-s
at a time instead of just a single one: if the iteration step does
not stop at the first good discriminant but will collect several
good ones. This way, we can select the input of the next step from
a set of new nodes.

The numbers have individual properties, which makes a difference
from the point of usability. The modified-ECPP predicts the
minimal value of $D$ which is still enough to produce at least one
new node for each $s_i$ produced by earlier steps and, building
upon this prediction, sets up a priority between them. It selects
the one with the highest priority as input for the next iteration
step. If the step does not provide output the limit $D$ will be
increased in order to use a new set of discriminants next time
when the node is selected. The priority is reevaluated after each
step because either there are new nodes or in case of no output
$D$ is increased. This way the possibility of getting stuck is
lower (details can be found in \cite{bo-ja-ki}).

\begin{sidewaystable}[]
\caption{The first rows of the proof of $n_1$}
\vspace{3mm}
\tiny
\begin{tabular}{| l | l | l | l | l | l | l |}
\hline \hline
$i$ & $s_i$ & $a_i$ & $b_i$ & $x_i$ & $y_i$ & $f_i$\\
\hline
1 & 165490139 & 148629518369919 & 154064198784106 & 1248188509129 &
156779851067219 & 1047222\\
\hline
\multirow {2}{*}
2 &
173304931274467 &
83727826741233492116 &
26748895837956005585 &
23717486180315890605 &
11528948633455951893 &
503211105\\
&
&
021 &
121 &
060 &
545 &\\
\hline
\multirow {2}{*}
3 &
87208965968598967476 &
27707400957247299977 &
38465375268196111041 &
28142518963869506523 &
20626644997054427792 &
6877885\\
&
679 &
4111947197 &
0774819991 &
604739355 &
7805954752 &\\
\hline
\multirow {2}{*}
4 &
59981323890093733168 &
44360713177559177572 &
11117040242765996352 &
19775111407066701253 &
18428910703962901519 &
408110\\
&
4100565579 &
547682547718474 &
2375118328811693 &
6508097973496758 &
3733563259937266 & \\
\hline
\multirow {2}{*}
5 &
24478978092786153542 &
18957568139328887813 &
10802087315828632193 &
69127839011266503431 &
35593578508679194394 &
4853\\
&
2029989890998131 &
3604105862968021099 &
95669478393483623141 &
9346422505776608550 &
5359772187301402486 &\\
\hline
\multirow {3}{*}
6 &
11879648068429120313 &
42083366473007987927 &
28055577648671991951 &
18066828358351637326 &
25827423550789508626 &
486045540848\\
&
59056929012033432579 &
62804820140778821039 &
75203213427185880693 &
30229471640748615572 &
77731097240769328473 &\\
&
&
92386685393 &
28257790262 &
09385839814 &
18909337972 &\\
\hline
\multirow {3}{*}
7 &
57740499705035302965 &
11380443233447151682 &
12206202132034258692 &
96784583503642168856 &
59710222084606095711 &
24\\
&
71991973651703959752 &
41036572010957871951 &
19783734272205524829 &
07646244567179724818 &
75786070086466618975 &\\
&
02732115661 &
7294768226496 &
2902663774895 &
972920682916 &
533228021487 &\\
\hline
\multirow {4}{*}
8 &
13857719929208472711 &
41694582424591111764 &
63775518208646414603 &
34124747439640525997 &
13433669109612418315 &
53096907911\\
&
77278058794700830584 &
43156453894775152544 &
03125686908495336379 &
72929688220778601290 &
59617778582791580192 &\\
&
4118454871693 &
45255867194500535484 &
05480470784855197824 &
76542652423518714847 &
42180614563041511515 &\\
&
&
548 &
527 &
638 &
824 &\\
\hline
\multirow {4}{*}
9 &
73580207893761171469 &
43072369720090946264 &
37871678106813894501 &
70859975965243426606 &
87503943548730935045 &
1271492\\
&
79557764345765335962 &
13350749758127925795 &
76380954770420227217 &
69499533970811791751 &
05661695045997191182 &\\
&
72772075969232836325 &
34831485143675218313 &
88363264525523890880 &
13479474574680812743 &
08741123870099436602 &\\
&
203 &
234009372 &
507401845 &
326336082 &
667558083 &\\
\hline
\multirow {4}{*}
{10} &
93556645695254179434 &
18992439425877599779 &
12661626283918399853 &
25459301193842798167 &
20234087314600234645 &
6762000366\\
&
47331860903525804841 &
81729792883189473382 &
21153195255459648921 &
48588095939934589673 &
56960079052485667673 &\\
&
40605445293133424460 &
67763448710029974361 &
78508965806686649574 &
54516398347837548526 &
33324851231828927752 &\\
&
870453037 &
4358086171211009277 &
2905390780807339518 &
9028646284881593862 &
7747596369915016860 &\\
\hline
\end{tabular}
\label{table:proof}
\end{sidewaystable}

\subsection{The proof}

On input $n$, a probable prime, the primality test results in a list, which provides sufficient data to prove the correctness of the sequence of the steps along the successful path. If we consider the length of the proof list as $\#L$, the $i^{th}$ list element, as the proof runs in reverse order, starting from the smallest $s_i$, corresponds to the $\#L-i^{th}$ step in the sequence and consists of $s_i$, $a_i$, $b_i$, $P_i$, $f_i$, where $s_if_i = m_i$ and $s_i$ is a probable prime, the factorization of $f_i$ is known, $y^2=x^3+a_ix+b_i$ is an elliptic curve of order $m_i$ over $\ZZ/s_{i+1}\ZZ$, and $P_i$ is a point on this curve that satisfies the condition $m_iP_i=0$, $f_iP_i \neq 0$. $P_i$ is given by its two coordinates $x_i$ and $y_i$. The correctness proof guarantees recursively that all $s_i$ are
genuine primes, and eventually that the input $n$ is prime.

Since the size of the above mentioned list is too large
(approximately 809 KB in txt form), the exact details can not be
presented in this paper. Instead of this, we give here only a
small part of this file (see in Table 1). The full text can be downloaded from page: 
\url{http://compalg.inf.elte.hu/tanszek/farkasg/proof-tri.txt}

\section*{Acknowledgements}

Prepared in the framework of application T\'AMOP 4.1.1/A-10/1/KONV-2010-0005
with the support of Universitas-Gy\H or Foundation (pages: 158--164);
The Project is supported by the European Union and co-financed by
the European Social Fund (grant agreement no. T\'AMOP
4.2.1/B-09/1/KMR-2010-0003) (pages: 165--171).

\bigskip
\rightline{\emph{Received:  June 11, 2011 {\tiny \raisebox{2pt}{$\bullet$\!}} Revised: September 14, 2011}} 


\begin{thebibliography}{99}

\bibitem{at-mo} A. O. L. \href{http://a-o-l-atkin.co.tv
}{Atkin}, F.
\href{http://www.lix.polytechnique.fr/~morain}{Morain}, Elliptic curves and 
\href{http://www.ams.org/journals/mcom/1993-61-203/S0025-5718-1993-1199989-X/S0025-5718-1993-1199989-X.pdf}
{primality proving}, \emph{Math. Comp.} \textbf{61,} 203 (1993) 29--68.


\bibitem{be} H. Belbachir, S. Bouroubi, A.
Khelladi,
\href{http://www.emis.de/journals/AMI/2008/ami2008-belbachir-bouroubi-abdelkader.pdf}
{Connection between ordinary multinomials}, Fibonacci numbers, Bell
polynomials and discrete uniform distribution, \emph{Ann. Math.
Inform.} \textbf{35} (2008) 21--30.


\bibitem{bo} B. A. Bondarenko, \emph{Generalized
Pascal Triangles and Pyramids, Their Fractals, Graphs and
Applications}, The \href{http://www.mscs.dal.ca/Fibonacci}{
Fibonacci} Association, Santa Clara, CA, USA, 1993.


\bibitem{bo-ja-ki}
W. \href{http://www.math.ru.nl/~bosma}{Bosma}, A.
\href{http://compalg.elte.hu/~ajarai}{J\'arai}, Gy.
\href{http://compalg.inf.elte.hu}{Kiss}, \textit{Better paths for
elliptic curve primality proofs},
\href{http://www.math.ru.nl/~bosma/pubs/reportfinal.pdf}{http://www.math.ru.nl/$\sim$bosma/pubs/reportfinal.pdf}, 2009.


\bibitem{magma}
W. \href{http://www.math.ru.nl/~bosma}{Bosma}, J. Cannon,
C. Playoust, The  \href{http://magma.maths.usyd.edu.au/magma}
{Magma} algebra system. I. The user language,
\emph{\href{http://www.sciencedirect.com/science/journal/07477171
}{J. Symbolic} Comput.}, \textbf{24}, 3-4 (1997) 235--265.



\bibitem{fa-ka} G. \href{http://compalg.inf.elte.hu}{Farkas}, G. \href{http://www.sze.hu/~kallos}{Kall\'os},
Prime numbers in \href{http://acta.maxwell.sze.hu/archive.html}{generalized Pascal triangles}, \emph{Acta Tech. Jaur.} \textbf{1,} 1 (2008) 109-118.






\bibitem{ka1} G. \href{http://www.sze.hu/~kallos}{Kall\'os}, \textit{A Pascal
h\'aromsz\"og \'altal\'anos\'\i t\'asai} (in Hungarian), Master
thesis, \href{http://www.elte.hu}{E\"otv\"os Lor\'and University},
Budapest, 1993.


\bibitem{ka2} G. \href{http://www.sze.hu/~kallos}{Kall\'os}, A generalization of  \href{http://www.numdam.org/numdam-bin/browse?id=AMBP_2006__13_1}
{Pascal's triangle}  using powers of base
numbers, \emph{Ann. Math. Blaise Pascal} \textbf{13,} 1 (2006)
1--15.


\bibitem{kn}
D. E.
\href{http://www-cs-faculty.stanford.edu/~knuth/}{Knuth},
\textsl{The Art of Computer Programming, Vol. 2.,}
\href{http://www.pearsonhighered.com/}{Addison-Wesley}, Reading, MA, USA, 1981.

\bibitem{ll} A. K. \href{http://www.win.tue.nl/~klenstra}
{Lenstra}, H. W. \href{http://math.berkeley.edu/~hwl}{Lenstra},
Algorithms in number theory, in: \emph{Handbook of Theoretical
Computer Science, Vol. A, Algorithms and Complexity,} Ed. J. van Leeuwen, Elsevier Science Publisher, B.V., Amsterdam; MIT Press, Cambridge, MA, 1990, pp. 673--715.


\bibitem{sl}
\textit{The On-line Encyclopedia of Integer Sequences}, Published
electronically at \href{http://oeis.org}{http://oeis.org}, 2011.



\end{thebibliography}
\end{document}